\documentclass[12pt]{article}
\usepackage{epstopdf,amsfonts,amsmath,amssymb}
\usepackage{graphicx}
\DeclareGraphicsExtensions{.eps}

\newcommand\encadremath[1]{\vbox{\hrule\hbox{\vrule\kern8pt
\vbox{\kern8pt \hbox{$\displaystyle #1$}\kern8pt}
\kern8pt\vrule}\hrule}}
\def\enca#1{\vbox{\hrule\hbox{
\vrule\kern8pt\vbox{\kern8pt \hbox{$\displaystyle #1$}
\kern8pt} \kern8pt\vrule}\hrule}}

\newcommand\framefig[1]{
\begin{figure}[bth]
\hrule\hbox{\vrule\kern8pt
\vbox{\kern8pt \vbox{
\begin{center}
{#1}
\end{center}
}\kern8pt}
\kern8pt\vrule}\hrule
\end{figure}
}

\newcommand\figureframex[3]{
\begin{figure}[bth]
\hrule\hbox{\vrule\kern8pt
\vbox{\kern8pt \vbox{
\begin{center}
{\mbox{\epsfxsize=#1.truecm\epsfbox{#2}}}
\end{center}
\caption{#3}
}\kern8pt}
\kern8pt\vrule}\hrule
\end{figure}
}
\newcommand\figureframey[3]{
\begin{figure}[bth]
\hrule\hbox{\vrule\kern8pt
\vbox{\kern8pt \vbox{
\begin{center}
{\mbox{\epsfysize=#1.truecm\epsfbox{#2}}}
\end{center}
\caption{#3}
}\kern8pt}
\kern8pt\vrule}\hrule
\end{figure}
}

\makeatletter
\@addtoreset{equation}{section}
\makeatother
\newtheorem{theorem}{Theorem}[section]

\newtheorem{remark}{Remark}[section]
\newtheorem{proposition}{Proposition}[section]
\newtheorem{lemma}{Lemma}[section]
\newtheorem{corollary}{Corollary}[section]
\newtheorem{definition}{Definition}[section]
\def\br{\begin{remark}\rm\small}
\def\er{\end{remark}}
\def\bt{\begin{theorem}}
\def\et{\end{theorem}}
\def\bd{\begin{definition}}
\def\ed{\end{definition}}
\def\bp{\begin{proposition}}
\def\ep{\end{proposition}}
\def\bl{\begin{lemma}}
\def\el{\end{lemma}}
\def\bc{\begin{corollary}}
\def\ec{\end{corollary}}
\def\beaq{\begin{eqnarray}}
\def\eeaq{\end{eqnarray}}
\newcommand{\proof}{{\noindent \bf proof:}$\quad$ }
\newcommand{\eproof}{ $\square$ }

\newcommand{\be}{\begin{equation}}
\newcommand{\ee}{\end{equation}}
\newcommand{\beq}{\begin{equation}}
\newcommand{\eeq}{\end{equation}}
\newcommand{\bea}{\begin{eqnarray}}
\newcommand{\eea}{\end{eqnarray}}

\newcommand{\e}{{\rm e}}
\newcommand{\ii}{{\rm i}\,}

\newcommand{\CC}{{\mathbb C}}

\newcommand{\ZZ}{{\mathbb Z}}

\newcommand{\spcurve}{{\cal S}}

\newcommand{\curve}{{\Sigma}}

\newcommand{\curverond}{\overset{\circ}{\curve}}

\newcommand{\acycle}{{\cal A}}
\newcommand{\bcycle}{{\cal B}}

\newcommand{\modsp}{{\mathcal M}}

\newcommand{\otimessym}{{\overset{\text{sym}}{\otimes} }}

\newcommand{\diag}{{\operatorname{diag}}}

\newcommand{\Ker}{{\rm Ker\ }}
\newcommand{\Img}{{\rm Im\ }}

\newcommand{\x}{{\rm x}}
\newcommand{\y}{{\rm y}}

\newcommand{\td}{\tilde}
\newcommand{\Res}{\mathop{\,\rm Res\,}}
\usepackage[pdftex]{hyperref}
\hypersetup{colorlinks,urlcolor=magenta,citecolor=red,linkcolor=blue,filecolor=black}

\begin{document}

\sloppy

\pagestyle{empty}
\hfill IPhT-T19/160 %,  IHES ??? %, CRM-???,
\addtolength{\baselineskip}{0.20\baselineskip}
\begin{center}
\vspace{26pt}
{\large \bf {
Topological Recursion,  Airy structures in the space of  cycles.
}}
\newline
\vspace{26pt}

{\sl B.\ Eynard}\hspace*{0.05cm}

\vspace{6pt}
Institut de Physique Th\'{e}orique de Saclay,\\
F-91191 Gif-sur-Yvette Cedex, France. CNRS URA 2306, F-91191 Gif-sur-Yvette, \\
CRM, Centre de recherches math\'ematiques  de Montr\'eal,\\
Universit\'e de Montr\'eal, QC, Canada.\\
IHES, Bures sur Yvette, France.
\end{center}
\vspace{20pt}
\begin{center}
{\bf Abstract:}
Topological recursion associates to a spectral curve, a sequence of meromorphic differential forms.
A tangent space to the "moduli space" of spectral curves (its space of deformations) is locally described by meromorphic 1-forms, and we use form-cycle duality  to re-express it in terms of cycles (generalized cycles).
This formulation allows to express the ABCD tensors of Quantum Airy Structures acting on the vector space of cycles, in an intrinsic spectral-curve geometric way.

\end{center}

%\end{center}

%-----------------------------ABSTRACT--------------------------------------
%
%Abstract

\vspace{0.5cm}
%\section*{}

%\newpage
\vspace{26pt}
\pagestyle{plain}
\setcounter{page}{1}

%*********************************************************************
%==================== ARTICLE =======================================%*********************************************************************

%\tableofcontents

\section{Introduction}

Topological recursion (T.R.) \cite{EO07} associates to a spectral curve (a plane curve with extra structure, see below), a sequence of mutilinear meromorphic differential forms, called the invariants of the spectral curve.
These invariants have many applications in mathematical physics, for example they compute the large $N$ asymptotic expansion of correlation functions in random matrix theory \cite{E04,CE05,EOreview}, and they also compute Gromov-Witten invariants in enumerative geometry \cite{BKMP,EO12,DOSS}, Hurwitz numbers \cite{BM07}, Weil-Petersson volumes \cite{EOwp}, Jones polynomials \cite{DFM,BEknots}, and many other interesting enumerative-algebro-geometric invariants happen to be the T.R. invariants of a suitable spectral curve \cite{EynICMreview}.

In \cite{KS17}, Kontsevich and Soibelman, then \cite{ABCD17}  reformulated T.R. in a geometric setting, instead of a spectral curve, their data was a vector space $V$ and its dual $V^*$, and tensors acting in them. They called it a quantum Airy structure.
In this setting, they showed that T.R. is equivalent to the statement that a family of quadratic differential operators built from the tensors, annihilate a wave function.

Here we shall re--translate KSABCD formalism in the language of spectral curves, and identify the vector space and operators.
The vector space $V^*$ should be a space of deformations of the Airy structure, and therefore it should be identified with the space of deformations of spectral curves, which in turn was shown to be isomorphic to a space of cycles \cite{EyTau17}.

\section{Spectral curves}

This section is a short reminder of some notions from \cite{EyTau17}.
We first define spectral curves, then forms and {\em generalized} cycles.

\subsection{Spectral curve}

\bd[Spectral curve]
The objects of the category %of spectral curves %$\SP$ 
are spectral-curves.

A spectral curve data is 
\beq
\spcurve=(\curve,\curverond,\x,\y,B)
\eeq
where
\begin{itemize}
\item $\curve$ is a smooth surface, not necessarily compact nor connected (it could be just a union of smooth discs, sometimes called a "local spectral curve")

\item $\curverond$ a Riemann surface, called the base, not necessarily compact nor connected

\item $\x:\curve\to\curverond$ a $C^\infty$ map. The pullback by $\x$ of the complex structure of $\curverond$, gives a complex structure to $\curve$, which is then seen as a Riemann surface (but remind that it's complex structure depends on the choice of $\x$).
$\curve$ is then a ramified cover of $\curverond$, and let $R$ the divisor of the ramification points, weighted by their order.

\item $\y$ is a meromorphic (with the above complex structure) 1-form on $\curve$.

\item $B$ is a symmetric meromorphic $1\otimes 1$ form on $\curve\times\curve$ (again with the above complex structure), with a double pole on the diagonal and no other pole, i.e $B\in H^0(\curve\times\curve, K_\curve\overset{\text{sym}}{\boxtimes} K_\curve(2\ \diag))$, normalized such that, in any chart
\beq
B(z_1,z_2) \sim \frac{dz_1 \otimes dz_2}{(z_1-z_2)^2} + \text{analytic at }z_1=z_2.
\eeq
\end{itemize}

A morphism 
\beq
(\curve,\curverond,\x,\y,B) \longrightarrow (\td\curve,\curverond,\td\x,\td\y,\td B)
\eeq
is a diffeomorphism $\phi:\curve\to\td\curve$, such that
$\td\x=\x\circ\phi$, and $\y=\phi^*\td\y$ and $B=\phi^*\td B$.

$\bullet$ if there is a morphism $\spcurve\to\td\spcurve$ and a morphism $\td\spcurve\to\spcurve$ that are inverse of each other, we say that $\spcurve$ and $\td\spcurve$ are isomorphic.

$\bullet$ a spectral curve $\spcurve$ is an equivalence class of spectral curve datas modulo isomorphisms.

Let us denote the moduli space of spectral curves modulo isomorphisms:
\beq
\modsp.
\eeq
%\beq
%\modsp=\SP/\text{diffeo}.
%\eeq

\ed
In all what follows, the base curve $\curverond$ is kept fixed, and in particular a local coordinate $x$ is chosen once for all in each chart.

%For the moment we don't put a topology.

\subsection{Bundle of meromorphic forms}

We have a vector bundle
\beq
\mathfrak M^1\to \modsp
\eeq
whose fiber is the infinite dimensional vector space
\beq
\mathfrak M^1(\spcurve) = \{\text{meromorphic 1-forms on } \curve\}.
\eeq
It is an infinite dimensional vector space, it is not countably generated.
%Notice that it depends only on $\curve,\curverond,\x$ in the spectral-curve data, it is independent of $\y$ and $B$.

\bd[Topological recursion]
Topological recursion (see \cite{EO07,EyTau17}) associates to a spectral curve a collection of forms as follows:
\beq
\omega_{0,1}(\spcurve)=\y
\qquad , \qquad 
\omega_{0,2}(\spcurve)=B,
\eeq
and for $2g-2+n>0$, we have that
\beq
\omega_{g,n}(\spcurve)\in H_0(\curve^n,K_\curve^{\boxtimes n,\text{ sym}}(*R))
\eeq
i.e. symmetric tensor products of $n$ 1-forms, with poles only over ramification points $R$. 
For readibility, we postpone the definition below in def.~\ref{defTR2}, def.~\ref{defTRhigher} where we will  introduce appropriate notations, or also the definition can be found in the literature \cite{EO07, EyTau17}.

For $n=0$, $\omega_{g,0}(\spcurve)$ is a 0-form i.e. a scalar and is denoted
\beq
\omega_{g,0}(\spcurve)=F_g(\spcurve) \in \CC.
\eeq

\ed

A property that will be useful to us is the homogeneity
\bp[Homogeneity]\label{prophomogeneity}
For $\lambda\in \CC^*$ define the rescaling of a spectral curve $\spcurve=(\curve,\curverond,\x,\y,B)$ :
\beq
\lambda \spcurve = (\curve,\curverond,\x,\lambda\y,B).
\eeq
We have \cite{EO07}
\beq
\omega_{g,n}(\lambda\spcurve) = \lambda^{2-2g-n}\omega_{g,n}(\spcurve).
\eeq
\ep

\subsection{Cycles}

Let $\spcurve$ a spectral curve with $(\curve,\curverond,\x,\y,B)$ a representent.

By Poincarr\'e duality, a cycle can be viewed as an element of the dual of $\mathfrak M^1(\spcurve)$, by the integration pairing:
\beq
<\gamma,\omega> = \int_\gamma \omega.
\eeq
Since $B$ is a $1\boxtimes 1$ form, integrating the second projection, produces a 1-form of the first projection, we define the 1-form $\hat B(\gamma)$ as:
\beq
\hat B(\gamma)(z_1) = \int_{z_2\in\gamma} B(z_1,z_2) .
\eeq
If $\gamma\in H_1(\curve,\ZZ)$, then $\hat B(\gamma)$ is a holomorphic 1-form on $\curve$.
However, we can also pair the 2nd projection in $B$ with any element of $\mathfrak M^1(\spcurve)^*$, the result will be a 1-form, but often this 1-form will not be meromorphic, it will not even be $C^\infty$, neither $C^0$.
We thus consider the subset of $\mathfrak M^1(\spcurve)^*$, for which the result of integrating $B$ yields a meromorphic 1-form, we call it the space of generalized cycles\footnote{This can be viewed as a Hodge decomposition of the dual space of meromorphic forms.}:

\bd[Generalized cycles]
\beq
\mathfrak M_1(\spcurve) = \{ \gamma\in \mathfrak M^1(\spcurve)^* \ | \ \hat B(\gamma)\in \mathfrak M^1(\spcurve) \}.
\eeq
By definition we have a map $\hat B: \mathfrak M_1(\spcurve) \to \mathfrak M^1(\spcurve)$.
It is proved in \cite{EyTau17} that this map is surjective, but not injective.
We have the  exact sequence
\beq
0 \to \Ker\hat B \to \mathfrak M_1(\spcurve) \overset{\hat B}{\to} \mathfrak M^1(\spcurve) \to 0.
\eeq
Notice that ordinary cycles are in the space of generalized cycles
\beq
H_1(\curve,\ZZ)\subset H_1(\curve,\CC) \subset \mathfrak M_1(\spcurve).
\eeq
It is customary to say that $\gamma$ is 
\begin{itemize}
\item a 1st kind cycle if $\hat B(\gamma)$ is a 1st kind form, i.e. has no poles, and thus 1st kind cycles are ordinary cycles  $\in H_1(\curve,\CC)$.
\item a 3rd kind cycle if $\hat B(\gamma)$ is a 3rd kind form, i.e. has at most simple poles. For example if $\gamma$ is a chain with boundary $\partial\gamma=D=\sum_i \alpha_i [p_i]$ then $\hat B(\gamma)$ is a 3rd kind cycle with simple poles $p_i$ of residues $\alpha_i$.

\item a 2nd kind cycle if $\hat B(\gamma)$ is a 2nd kind form, i.e. has some poles of degree $\geq 2$.

\end{itemize}

\ed

\bd[Intersection and symplectic structure]
We define the intersection of generalized cycles as
\beq
\gamma_1 \cap \gamma_2 = 
\frac{1}{2\pi\ii} \left( \int_{\gamma_1}\hat B(\gamma_2) - \int_{\gamma_2}\hat B(\gamma_1) \right).
\eeq
It is proved in \cite{EyTau17} that the intersection defines a non--degenerate symplectic form on $\mathfrak M_1(\spcurve)$.
This intersection matches the usual intersection on  $H_1(\curve,\CC)$.

\ed

\bd[Cycle bundle]
The infinite dimensional vector bundle
\beq
\mathfrak M_1\to\modsp
\eeq
whose fiber is the space of generalized cycles, is a flat bundle.
It admits a flat connection \cite{EyTau17}.
\ed
The flat connection is somehow a pullback of cycles from the base curve $\curverond$, tensored with pullbacks of local meromorphic functions on neighborhood of the cycle in $\curverond$.
Since it depends only on $\curverond$, it is flat (cf the Gauss-Manin connection).

The map $\hat B$ that sends cycles to forms, is not invertible, it has a huge kernel ($\Ker \hat B$ is a Lagrangian of $\mathfrak M_1(\spcurve)$), however there is a right inverse of $\hat B$ as follows.

\bd[forms to cycles \cite{EyTau17}]
There is a linear map
\bea
\check B: \mathfrak M^1(\spcurve) \to \mathfrak M_1(\spcurve),
\eea
such that
\beq
\hat B \circ \check B = \text{Id}
\qquad , \quad 
\check B \circ \hat B = \Pi \qquad \Pi^2=\Pi.
\eeq
$\Pi$ is the projection onto $\Img \check B$, parallel to $\Ker\hat B$.

$\Ker \hat B$ and $\Img \check B$ are Lagrangian submanifolds, and
\beq
\mathfrak M_1(\spcurve) = \Ker\hat B \oplus \Img \check B.
\eeq

\ed
We leave the reader see the actual definition of $\check B$ in \cite{EyTau17}, but to get an idea, imagine that we would have $\{\acycle_i\}_{i\in I}$ a basis of $\mathfrak M_1(\spcurve)$, and $I_{i,j} = \acycle_i \cap \acycle_j$ its intersection matrix,
the definition of $\check B(\omega) $ for a meromorphic 1-form $\omega$ would be
\beq\label{eqdefChat}
\check B(\omega) =\frac{1}{2\pi\ii} \sum_{i,j} \left(  (I^{-1})_{i,j} \int_{\acycle_j}\omega \right) \ \acycle_i.
\eeq

In fact $\dim \mathfrak M_1(\spcurve)=\infty$ and is not countably generated, so it seems that this definition involving infinite sums would be ill-defined. 
The actual geometric definition, given in \cite{EyTau17} is based on the Riemann-bilinear identity. However, \eqref{eqdefChat} is morally correct and can be used in practice. Indeed \eqref{eqdefChat} is invariant under change of basis, and for a given meromorphic 1-form $\omega$, it is always possible to find a basis in which only finitely many terms in \eqref{eqdefChat} are non-vanishing, or in which the sum is absolutely convergent.
The true definition of \cite{EyTau17} will however  not be needed in the rest of this article.

\subsection{Tangent moduli space}

In \cite{EyTau17} it was shown that the tangent space $T_\spcurve \modsp$ (i.e. the space of deformations of spectral curves) is in fact isomorphic to a space of meromorphic forms, and thanks to the dualities above, it can be embedded in a space of cycles:
\beq
T_\spcurve \modsp \hookrightarrow \mathfrak M^1(\spcurve) \oplus (\mathfrak M^1(\spcurve)\overset{\text{sym}}{\otimes} \mathfrak M^1(\spcurve)),
\eeq
with the following  map:
\bd[cycles to tangent vectors \cite{EyTau17}]
We define the map $\partial: \mathfrak M_1(\spcurve) \oplus (\mathfrak M_1(\spcurve)\overset{\text{sym}}{\otimes} \mathfrak M_1(\spcurve)) \to T_\spcurve \modsp$ as follows:

for $\gamma \in \mathfrak M_1(\spcurve)$:
\bea
\partial_\gamma \x & =& 0 \cr
\partial_\gamma \y & =& \hat B(\gamma) \cr
\partial_\gamma B(z_1,z_2) & = & \int_{z\in \gamma} \omega_{0,3}(\spcurve;z_1,z_2,z) \cr
\eea
and for ${\gamma_1\otimessym \gamma_2}=\frac12(\gamma_1\otimes \gamma_2+\gamma_2\otimes \gamma_1) \in  \mathfrak M_1(\spcurve)\overset{\text{sym}}{\otimes} \mathfrak M_1(\spcurve)$:
\bea
\partial_{\gamma_1\otimessym \gamma_2} \x & =& 0 \cr
\partial_{\gamma_1\otimessym \gamma_2} \y & = & 0 \cr
\partial_{\gamma_1\otimessym \gamma_2} B 
& =& \frac12 \left( \hat B(\gamma_1) \otimes \hat B(\gamma_2) + \hat B(\gamma_2) \otimes \hat B(\gamma_1) \right) \cr
\eea

\ed
Remark that the mapping between tangent space and cycles is very similar to the Goldman cycles, and indeed the intersection symplectic form, is mapped to the Goldman bracket, see \cite{Goldman, BelliardEynard2019}.

The following theorem \cite{EyTau17} follows immediately from \cite{EO07}
\bt
For $(g,n)\neq (0,0)$
\beq\label{eqdgammaomgn}
\partial_\gamma \omega_{g,n} = \int_\gamma \omega_{g,n+1}
\eeq

\bea
\partial_{\gamma_1\otimessym \gamma_2} \omega_{g,n} (z_1,\dots,z_n)
&=& \int_{\gamma_1}\int_{\gamma_2} \Big( \omega_{g-1,n+2}(z_1,\dots,z_n,z,z') \cr
&&  + \sum_{g_1+g_2=g, I_1\sqcup I_2=\{z_1,\dots,z_n\}} \omega_{g_1,1+|I_1|}(I_1,z)\omega_{g_2,1+|I_2|}(I_2,z') ) \Big) \cr
\eea

\et

%The reason why \eqref{eqdgammaomgn} fails for $(g,n)=(0,0)$ is that
%\beq
%\partial_{\gamma'}(\int_{\gamma}\omega_{0,1})-\partial_\gamma(\int_{\gamma'}\omega_{0,1}) = 2\pi\ii \gamma\cap \gamma'
%\eeq
%i.e. there is a curvature equal to the intersection, which prevents from defining $\omega_{0,0}(\spcurve)$.
%In fact $\omega_{0,0}(\spcurve)$ can be defined only after choosing a Lagrangian, on which the curvature vanishes. See \cite{EynTau17}.

\subsection{Hirota derivative}

\bd[Hirota derivative]
Let $z\in \curve$ a generic point, and let $\bcycle_{z,1}\in \mathfrak M_1(\spcurve)$ be the linear form defined by $\int_{\bcycle_{z,1}} \omega = \Res_{p\to z} \frac{1}{\x(p)-\x(z)} \omega$.
We define
\beq
\Delta_z = d\x(z) \otimes \partial_{\bcycle{z,1}}.
\eeq
It is the tensor product of a 1-form by a tangent vector
\beq
\Delta \in H^0(\curve,K_{\curve}\otimes T_{\spcurve}\modsp).
\eeq

\ed

We have $\partial_\gamma=<\gamma,\Delta>$, in the sense that if $f$ is a function  on $\modsp$ we have
\beq
\partial_\gamma f = \int_\gamma \Delta f.
\eeq
%If $\gamma\in \Ker\hat B$ we have
%\beq
%\gamma\in\Ker\hat B \ \ \implies \ \ \int_\gamma \Delta f=0.
%\eeq

It is called the insertion operator because of the following property:
\bt[Insertion operator \cite{EO07}]
For $(g,n)\neq (0,0)$
\beq
\Delta_z \omega_{g,n}(z_1,\dots,z_n) = \omega_{g,n+1}(z,z_1,\dots,z_n).
\eeq
\et

%
%\subsection{Integer cycles}
%
%In \cite{EyTau17}, it was shown that $\mathfrak M_1(\spcurve)$ contains a lattice
%\beq
%\mathfrak M_1(\spcurve,\ZZ) \subset \mathfrak M_1(\spcurve).
%\eeq
%An explicit integer Darboux basis was constructed (and is not unique):
%\beq
%\acycle_i \cap \bcycle_j=\delta_{i,j}.
%\eeq
%This shows that the bundle
%\beq
%\mathfrak M^1 \to \modsp
%\quad , \quad \text{fiber}=\mathfrak M^1(\spcurve).
%\eeq
%is a flat bundle, that we can equip with the trivial connection in an integer Darboux basis.

%
%\subsection{Tautological cycle}
% 
%????????
%
%useful ?
%
%??????? 
% 
% 
% 
%\bd[Tautological cycle]
%\beq
%\hat \Gamma = \check B(\y).
%\eeq
%
%\ed
%
%We recall that in a basis we have
%\beq
%\hat\Gamma=\frac{1}{2\pi\ii} \sum_{i,j} \left(  (I^{-1})_{i,j} \int_{\acycle_j}\y \right) \ \acycle_i.
%\eeq
%
%The tautological cycle is the dual of the tautological 1-form $\y$.
%It defines a section of the bundle $\mathfrak M_1\to \modsp$, this is not a flat section.
%Instead we have
%\beq
%\partial_\gamma \hat\Gamma = \Pi(\gamma)=\check B(\hat B(\gamma)).
%\eeq
%
%\bt[Dilaton \cite{EO07}]
%\beq
%\partial_{\hat\Gamma} \omega_{g,n} = (2-2g-n) \omega_{g,n}.
%\eeq
%\et

\section{Quantum Airy structures}

\subsection{Topological recursion}

Let $\mathcal L$ a flat Lagrangian in the total space of the bundle $\mathfrak M_1\to\modsp $, we assume that it is generically transverse to $\Ker\hat B$.
Let $\mathcal S$ a spectral curve and $V=\mathcal L\cap \text{fiber}(\spcurve)$.
$V$ is an infinite dimensional vector space, it is a Lagrangian in the fiber $\mathfrak M_1(\spcurve)$.
We assume  (generically true) 
\beq
\mathfrak M_1(\spcurve) = V\oplus \Ker\hat B.
\eeq
We have the map 
\beq
\hat B: V\to \mathfrak M^1(\spcurve) ,
\eeq
which is an isomorphism.
Moreover  the inclusion map $\mathfrak M_1(\spcurve)\subset \mathfrak M^1(\spcurve)^*$ restricts to an isomorphism $V^*\sim \mathfrak M^1(\spcurve)$ so that $\hat B$ can be viewed as the dualizing map
\beq
\hat B: V \to V^*.
\eeq
We also have the Hirota operator $\Delta\in \mathfrak M^1(\spcurve)\otimes T_{\spcurve}\modsp$.

We shall use the vector space $V$ and its dual $V^*$ as the vector space of the Airy structure of \cite{KS17}, and we shall define some tensors acting on them.
% on the vector space $V$ and its dual $V^*\sim \mathfrak M^1(\spcurve)$.

First, observe that the projection $\x:\curve\to\curverond$ has ramification points, let $R=\sum_a (\operatorname{order}_a (\x-\x(a))-1).a$ the divisor of ramification points.
In this section assume that all ramification points are generic, of order  $r_a=\operatorname{order}_a (\x-\x(a))=2$.
We postpone higher order cases to section \ref{sec:higherorderbp}.

There are 2 sheets meeting at $a$. Let $\sigma_a\neq \text{Id}$ the unique holomorphic involution  in a simply connected neighborhood of $a$ exchanging the 2 sheets, i.e. such that $\x(\sigma_a(z))=\x(z)$ and $\sigma_a(a)=a$.
If $\omega$ is a 1-form holomorphic in a simply connected neighborhood of $a$, we define $d_a^{-1} \omega = \int_{z'=a}^z \omega(z')$, the unique primitive of $\omega$ that vanishes at $z=a$.

\bd[Recursion kernels \cite{EO07}]

Define $K_2: V^*\otimes V^*\to V^*$
\beq
K_2(\omega,\omega')
= - \sum_{a\in R} \Res_{a} \frac{d_a^{-1} B}{\y-\sigma_a^*\y} \omega \otimes \sigma_a^*\omega',
\eeq
where the $d_a^{-1}$ and the residue act only on the second projection of $B$, i.e. $K_2 $ returns a 1-form in the 1st projection of $B$.
In formulas \cite{EO07} this reads
\beq
K_2(\omega,\omega')(z_1)
= - \sum_{a\in R} \Res_{z_2\to a} \frac{\int_{z'=a}^{z_2} B(z_1,z')}{\y(z_2)-\y(\sigma_a(z_2))} \omega(z_2) \otimes \omega'(\sigma_a(z_2)).
\eeq
\ed

\bd[Topological recursion \cite{EO07}]\label{defTR2}
The $\omega_{g,n}\in (V^*)^{\otimes n}$ are defined by
\beq
\omega_{0,1}=\y,
\quad , \quad
\omega_{0,2}=B,
\eeq
and by the recursion for $2g-2+n+1>0$ and $n\geq 0$
\bea\label{eqdefTRK2}
\omega_{g,n+1}(z_0,z_1,\dots,z_n) 
&=& K_2\Big(  \omega_{g-1,n+2}(.,.,z_1,\dots,z_n) \cr
&& + \sum_{g_1+g_2=g, I_1\sqcup I_2}^{\text{no }(0,1)} \omega_{g_1,n_1+1}(.,I_1) \otimes \omega_{g_2,n_2+1}(.,I_2) \Big)
\eea
where $K_2$ acts on the dots variables.
and for $g\geq 2$
\beq
\omega_{g,0}=F_g = \frac{1}{2-2g} <\hat\eta,\omega_{g,1}>
\eeq

We leave the reader look the definition of $F_1=\omega_{1,0}$ in \cite{EO07, KoKo}, and $F_0$ will not be needed here.
\ed

\bt[\cite{CE05,EO07}]

We have
\beq
\partial_\gamma K_2(\omega,\omega') 
= K_2(\hat B(\gamma),K_2(\omega,\omega'))
+ K_2(\partial_\gamma\omega,\omega') + K_2(\omega,\partial_\gamma\omega').
\eeq

\et

\subsection{Tensors and Airy structure}

Following KS \cite{KS17}, we define
\bd[ABCD]

We define the following tensors
\begin{itemize}

\item $A=\omega_{0,3}\in (V^*\otimes V^*\otimes V^*)^{\text sym}$, i.e.
\beq
A(\gamma_1,\gamma_2,\gamma_3) = \int_{\gamma_1}\int_{\gamma_2}\int_{\gamma_3} \omega_{0,3}.
\eeq

\item $D=\omega_{1,1}\in V^*$, i.e.
\beq
D(\gamma) = \int_{\gamma} \omega_{1,1}.
\eeq

\item $C\in V^*\otimes (V\otimes V)^{\text sym}$, is the dual of the recursion kernel, $C=2K_2^* $:
\beq
C(\gamma,\omega,\omega')
=2 \int_\gamma K_2(\omega,\omega').
\eeq

\item and composing with $\hat B$ we define $B\in V^*\otimes V^*\otimes V$:
\beq
B(\gamma_1,\gamma_2,\omega) = 2C(\gamma_1,\hat B(\gamma_2),\omega) .
\eeq

\end{itemize}
Here in the context of topological recursion for spectral curves we have the relations
\beq
A(\gamma_1,\gamma_2,\gamma_3) =  C(\gamma_1,\hat B(\gamma_2),\hat B(\gamma_3)),
\eeq
which we write
\beq
B=2C^{.*.} 
\qquad , \qquad
A= C^{.**},
\eeq
and notice that $\omega_{0,2} \in V^*\otimes V^*$, and  $D$ can be written 
\beq
D(\gamma) = C(\gamma,\omega_{0,2}).
\eeq

\ed

These tensors are such that
\beq
\omega_{0,3}=A
\eeq
\beq
\omega_{1,1}=D
\eeq
and $\omega_{g,n}$ with $2g-2+n>1$ is given by recursively applying the tensors $C$ and $B$
\bea
2\int_{z_0\in \gamma} \omega_{g,n+1}(z_0,z_1,\dots,z_n) &=&  C(\gamma,\omega_{g-1,n+2}(.,.,z_1,\dots,z_n)) \cr
&& + \sum^{\text{stable}}_{g_1+g_2=g, \ I_1\sqcup I_2=\{z_1,\dots,z_n\}} C(\gamma,\omega_{g_1,1+|I_1|}(.,I_1),\omega_{g_2,1+|I_2|}(.,I_2))\cr
&& + 2\sum_{j=1}^n B(\gamma,\omega_{0,2}(.,z_j),\omega_{g,n}(..,\hat z_j))
\eea
where the dot--variables are the ones on which the tensors act, and $\hat z_j$ means $\{z_1,\dots,z_n\}\setminus z_j$, and stable means that we exclude $(g_i,n_i)=(0,1),(0,2)$ from the sum.

These are thus the tensors of the Quantum Airy Structure of \cite{KS17,ABCD17}.

%\bp[Quantum Airy Structure]
%%They satisfy the HIX relation of \cite{ABCD17}.
%\ep

\subsection{Wave function}

\bd[Tautological cycle]
We define the tautological cycle as the dual of $\omega_{0,1}$ in $V$, i.e. 
\beq
\hat\eta=\omega_{0,1}^*=\Pi_V^{\parallel\Ker\hat B} \check B(\omega_{0,1}).
\eeq
where $\Pi_V^{\parallel\Ker\hat B} $ is the projection on $V$ parallel to $\Ker\hat B$.
\ed
Some of its properties are:
\beq
\forall \gamma\in V , \qquad \partial_\gamma \hat\eta = \gamma,
\eeq
\beq
\hat B(\hat\eta) = \omega_{0,1},
\eeq
and the dilaton equation \cite{EO07} amounts to
\beq
\forall \ 2-2g-n<0 \ , \qquad \partial_{\hat\eta} \omega_{g,n} = (2-2g-n) \omega_{g,n}.
\eeq

\bd
We define $F_0$ as
\beq
F_{0,\mathcal L}(\spcurve) = \frac12 <\hat \eta,\omega_{0,1}>.
\eeq
Notice that $F_0$ depends on our choice of Lagrangian $\mathcal L$.
\ed

Remark that if we wouldn't project $\check B(\omega_{0,1})$ on $V$, we would have
\beq
<\check B(\omega_{0,1}),\omega_{0,1}>=0.
\eeq

%
%\bd[Partition function]
%
%Let
%
%
%\beq
%\ln Z_{\mathcal L}(\hbar^{-1}\spcurve)=\hbar^{-2}F_{0,\mathcal L}(\spcurve) + \sum_{g\geq 1} \hbar^{2g-2} F_g(\spcurve).
%\eeq
%
%\ed
%
%
%?????????????
%
%Notice that
%\beq
%\mathcal F(\mathcal L)= \sum_{g,n} (-1)^n \frac{\hbar^{2g-2+n}}{n!} <\hat\eta\otimes \dots \otimes \hat\eta, \omega_{g,n}>
%\eeq
%is constant in $\mathcal L$, i.e. $\forall \gamma\in V$
%\beq
%\partial_\gamma \mathcal F=0.
%\eeq
%So we have
%\beq
%Z(\hbar^{-1}\spcurve-\hbar^{-1}\hat\eta) = e^{\mathcal F(\mathcal L)}
%\eeq

\bd[Wave function]\label{defwavef}
Define for $\gamma'\in V$:
\bea
Z'(\hbar^{-1}\spcurve,\gamma') 
%&=& Z_{\mathcal L}(\hbar^{-1}\spcurve+\gamma')e^{-\hbar^{-2}F_{0,\mathcal L}(\spcurve)} \cr
&=& e^{\sum{(g,n)\neq (0,0)} \frac{\hbar^{2g-2+n}}{n!} <\gamma'^{\otimes n},\omega_{g,n}(\spcurve)>} .
\eea
It is defined as a formal power series of $\hbar$, namely
\beq
\hbar\ln Z' \in \CC[[\hbar]].
\eeq
and all equations we are going to write from now on, are understood in $\CC[[\hbar]]$.

\ed

The notation $\hbar^{-1}\spcurve$ comes from the homogeneity prop.~\ref{prophomogeneity}.

Remark that formally we have the Sato formula: wave function = shifted partition function, i.e.
\beq
Z'(\hbar^{-1}\spcurve,\gamma') 
= Z(\hbar^{-1}\spcurve+\gamma')  \ e^{-\hbar^{-2}F_0(\spcurve)}
= Z(e^{\partial_{\gamma'}}\hbar^{-1}\spcurve) \ e^{-\hbar^{-2}F_0(\spcurve)}
\eeq
where $e^{\partial_{\gamma'}}$ is the exponential of the flow of the tangent vector $\partial_{\gamma'}$, and 
\beq
Z(\hbar^{-1}\spcurve) =  \e^{\sum_{g=0}^\infty \hbar^{2g-2} F_g(\spcurve) }.
\eeq
In the ratio $Z'$, we see that $F_0$ cancels.

\subsection{Quadratic PDE}

Let us revisit Kontsevich-Soibelman in this language.

\bd
Define the quadratic differential operators
\beq\label{defoperatorK2}
L = \hbar\Delta-\hat B(\gamma') - \hbar 
\left( \omega_{1,1}+ \hbar^2 K_2(\Delta\otimes\Delta)\right)
\eeq
and for any $\gamma\in V$
\beq
L_\gamma = <\gamma,L> = \hbar\partial_\gamma - \int_\gamma \hat B(\gamma') -\hbar 
\left(  D(\gamma) + \frac{\hbar^2}2 C(\gamma,\Delta\otimes\Delta)\right)
\eeq

\ed
Remark that if we assign an "order" $\hbar^{-1}$ to derivatives, we see that $L$ is a $O(\hbar)$ deformation of the Hirota operator shifted by  $\hat B(\gamma')$:
\beq
L=\hbar\Delta-\hat B(\gamma') + O(\hbar).
\eeq

The following is the main theorem:

\bt[Annihilating the wave function ]\label{mainth2}

%\beq
%L_\gamma = \hbar\partial_\gamma-\hbar D(\gamma)-\frac{\hbar}2 (\hbar^2 C(\gamma,\Delta\otimes \Delta) + 2\hbar B(\gamma,\gamma',\Delta)+A(\gamma,\gamma',\gamma')) .
%\eeq
We have $\forall \gamma'\in V$
\beq\label{maineq2}
L .Z'(\hbar^{-1}\spcurve,\gamma')=0
\eeq
or equivalently $\forall \gamma\in V$
\beq
L_\gamma .Z'(\hbar^{-1}\spcurve,\gamma')=0.
\eeq

\et

\proof
This is merely a way of rewriting topological recursion eq~\ref{eqdefTRK2}.

\bea
\Delta \ln Z'
&=& \hbar^{-1}\hat B(\gamma') + \sum_{2g-2+n+1>0} \frac{\hbar^{2g-2+n}}{n!} <\gamma'^{\otimes n},\omega_{g,n+1}> \cr
&=& \hbar^{-1}\hat B(\gamma') 
+  \sum_{2g-2+n+1>0} \frac{\hbar^{2g-2+n}}{n!} K_2(<\gamma'^{\otimes n},\omega_{g-1,n+2}>) \cr
&& +  \sum_{2g-2+n+1>0} \sum_{g_1+g_2=g} \sum_{n_1+n_2=n} \frac{\hbar^{2g-2+n}}{n_1! n_2!} K_2(<\gamma'^{\otimes n_1},\omega_{g_1,n_1+1}>,<\gamma'^{\otimes n_2},\omega_{g_2,n_2+1}>)
\cr
&=& \hbar^{-1}\hat B(\gamma') +\omega_{1,1} + \hbar^2 K_2(\Delta^2\ln Z') + \hbar^2 K_2(\Delta\ln Z',\Delta\ln Z') \cr
&=& \hbar^{-1}\hat B(\gamma') +\omega_{1,1} + \hbar^2 K_2(Z'^{-1}\Delta^2  Z') 
\eea

\eproof

\subsection{Higher order branchpoints}
\label{sec:higherorderbp}

At a ramification point $a$ of order $r_a\geq 2$, there are $r_a$ sheets meeting, $x$ is locally like $x(z)\sim x(a)+c_a z^{r_a}$, and there is a local Galois group
\beq
G_a=\ZZ_{r_a}
\eeq
exchanging the sheets, acting by multiplication of $z$ by a $r_a$th root of unity.
Let $G_a^*=G_a\setminus\{\text{Id}\}$ (for $r_a=2$, we recognize $G_a^*=\{\sigma_a\}$ the local involution.)
For $k\geq 2$, we shall consider the set of all possible $(k-1)-$uples of $G_a^*$, denoted $\sigma=(\sigma_2,\sigma_3,\dots,\sigma_k)$.
If $k>r_a$ this set is empty.

\bd
For $k\geq 2$, define $K_k:V^{\otimes k}\to V^*$
\beq
K_k(\omega_1,\dots,\omega_k)
= - \sum_a  \sum_{\sigma\subset_k G_a^*} \Res_{a} \frac{d_a^{-1} B}{\prod_{i=2}^k (\y-\sigma_i^*\y)} \omega_1 \otimes \sigma_2^*\omega_2 \otimes \dots \otimes \sigma_k^*\omega_k,
\eeq
and after dualizing, define $C_k=K_k^*\in V^*\otimes V^{\otimes k}$:
\beq
C_k(\gamma,\omega_1,\dots,\omega_k)=\int_\gamma K_k(\omega_1,\dots,\omega_k)
\eeq
\ed

\bd[Topological recursion \cite{BouEyn13}]
\label{defTRhigher}

For $2g-2+n+1> 0$:
\bea
\omega_{g,n+1}(z_0,z_1,\dots,z_n) 
&=& \sum_{k=2}^{\max r_a} \sum_{\mu \vdash k} 
\sum_{J_1\sqcup \dots \sqcup J_{\ell(\mu)}=\{z_1,\dots,z_n\}}
\sum_{g_1,\dots,g_{\ell(\mu)}, \ \sum_{i}g_i=g-k+\ell(\mu)} \cr
&& K_k\Big(  \omega_{g_i,|\mu_i|+|J_i|}(J_i,\mu_i) \Big)
\eea
where $K_k$ acts on the $\mu_i$ variables.
and for $g\geq 2$
\beq
\omega_{g,0}=F_g = \frac{1}{2-2g} <\hat\eta,\omega_{g,1}>
\eeq

\ed

\bt
The following differential operator (whose order is $\max_{a\in R} r_a$)
\beq
L=\hbar\Delta-\hbar \sum_{k\geq 2}  \sum_{l=0}^k \frac{k!}{l!(k-l)!} K_k(\hbar^l \Delta^{\otimes l}, U_{k-l} )
\eeq
where $U_k$ is given by
\beq
U_k = e^{-F_0} \Delta^{k, \text{ no discs}} e^{F^0}= \left( e^{-F_0} \Delta^{k} e^{F^0}\right)_{\omega_{0,1}\to 0}.
\eeq
\beq
U_1=0
\quad , \quad 
U_2=\omega_{0,2}
\quad , \quad 
U_3 = \omega_{0,3}
\quad , \quad 
U_4 = \omega_{0,4} + 3 \omega_{0,2}\omega_{0,2} \ , \dots
\eeq
annihiliates $Z'$
\beq\label{maineqk}
L.Z'=0.
\eeq

\et

\proof
Again this is a mere rewriting of topological recursion.
\bea
\Delta \ln Z'
&=& \hat B(\gamma') + \sum_k \sum_{\mu, \ |\mu|=k} c_\mu K_k(\prod_i \Delta^{\mu_i}\ln Z' + \delta_{\mu_i\geq 2} \Delta^{\mu_i}F_0)
\eea

\bea
\Delta Z' 
&=&  \hat B(\gamma')Z' +   \sum_k \sum_{a=0}^k \frac{k!}{a! (k-a)!} K_k(\Delta^{k-a} Z', e^{-F_0} \Delta^{a, \text{ no discs}}e^{F_0}) \cr
\eea
%FAUX

\eproof

Example with $r=3$ (and written with $\hbar=1$):
\bea
\Delta \ln Z'
&=& \hat B(\gamma') \cr
&& + K_2(\Delta^2\ln Z' + \Delta^{2}F_0) + K_2(\Delta\ln Z', \Delta \ln Z') \cr
&& + K_3(\Delta^{3}\ln Z' + \Delta^{3}F_0) \cr
&& +3 K_3(\Delta^{2}\ln Z' + \Delta^{2}F_0,\Delta \ln Z') \cr
&& + K_3(\Delta\ln Z' ,\Delta \ln Z',\Delta \ln Z') \cr
&=& \hat B(\gamma') + \omega_{1,1} + K_3(\omega_{0,3}) \cr
&& + K_2(\Delta^2\ln Z' ) + K_2(\Delta\ln Z', \Delta \ln Z') \cr
&& + K_3(\Delta^{3}\ln Z' ) \cr
&& +3 K_3(\Delta^{2}\ln Z',\Delta \ln Z') \cr
&& +3 K_3( \omega_{0,2},\Delta \ln Z') \cr
&& + K_3(\Delta\ln Z' ,\Delta \ln Z',\Delta \ln Z') \cr
\eea
i.e. the operator that annihilates $Z'$ is 
\beq
L = \Delta - \hat B(\gamma') - \omega_{1,1}- K_2(\Delta^2)-3 K_3(\omega_{0,2}\otimes\Delta)-K_3(\Delta^3).
\eeq

\subsection{Local times formulation}

Let us show how to recover the usual KS Airy-structures formulation  \cite{KS17,ABCD17}.
This is in some sense a mere "change of basis" in the space of cycles, however many subtleties arise because the local cycles, depending on the branchpoints $x(a)$ are not flat sections of the cycle-bundle. The flat connection is not trivial in that non-flat  basis.

\subsubsection{Local cycles and local times}

In order to describe the $\omega_{g,n}$, which have poles at ramification points, we introduce the following family of cycles (called local cycles).
Let $a$ a ramification point of order $r_a$, and let the generalized cycle $\Gamma_{a,k}\in \mathfrak M_1(\spcurve)$ be defined as a a linear form $\in \mathfrak M^1(\spcurve)^*$, which acts on any meromorphic 1-form $\omega$ as
\beq
\int_{\Gamma_{a,k}}\omega = <\Gamma_{a,k},\omega> = \Res_{z\to a}  (x-x(a))^{-k/r_a} \omega(z).
\eeq
%$\mathcal C_a$ an "infinitely small" clockwise circle around $a$) and let
%\beq
%\Gamma_{a,k} = \frac{1}{2\pi\ii } (x-x(a))^{-k/r_a} \mathcal C_a,
%\eeq
%\beq
%\bcycle_{a,k} = \frac{1}{2\pi\ii k} (x-x(a))^{k/r_a} \mathcal C_a.
%\eeq
They intersect as
\beq
\Gamma_{a,k}\cap \Gamma_{b,j} = \frac{ |k|}{2\pi\ii} \ \delta_{a,b} \ \delta_{k,-j}. 
\eeq

Define the local times for $k\geq 1$ (in fact they are 0 if $k\leq 0$)
\beq
t_{a,k} = \oint_{\Gamma_{a,k}} \omega_{0,1}
\eeq
they are the coefficients of the Taylor expansion of $\omega_{0,1}$ at $a$
\beq
\omega_{0,1} \sim \frac{1}{r_a}\sum_{k=1}^\infty t_{a,k} (x(z)-x(a))^{\frac{k-r_a}{r_a}}dx(z)  
\ + \text{analytic at }a.
\eeq
It is important to keep in mind that ramification points move when we deform the spectral curve, and these cycles are not flat.
We have
\beq
\Delta x(a) = -\frac{r_a}{t_{a,r+1}} \hat B(\Gamma_{a,1})
\eeq
\beq
\Delta \Gamma_{a,k} = -\frac{k}{t_{a,r+1}}  \hat B(\Gamma_{a,1})\otimes \Gamma_{a,k+r_a}. 
\eeq
%\beq
%\Delta \bcycle_{a,k} = -\frac{k-r_a}{t_{a,r+1}}  \hat B(\acycle_{a,1})\otimes \bcycle_{a,k-r_a}
%\eeq
%(and if $k<r_a$, replace $(k-r_a)\bcycle_{a,k-r_a}$ with $-\acycle_{a,r_a-k}$.)

For $k\geq 1$, these local times are not flat coordinates:
\beq
\Delta t_{a,k} =  \left( \hat B(\Gamma_{a,k}) - k \frac{t_{a,r+k}}{t_{a,r+1}} \hat B(\Gamma_{a,1}) \right).
\eeq

%
%\beq 
%\partial_{\bcycle_{a,k}} = \frac{\partial}{\partial t_{a,k}}. ????
%\eeq

%\beq
%\Delta_z \sim \sum_{k=1}^\infty \frac{k}{r_a} (x(z)-x(a))^{-\frac{k+r_a}{r_a}}dx(z) \frac{\partial}{\partial t_{a,k}} \quad + \text{analytic at }a. ??????
%\eeq

\subsubsection{Wave function}

In def.~\ref{defwavef}, let us choose a cycle $\gamma'$, written
\beq
\Gamma' = \sum_{a} \sum_{k\geq r_a+1} \frac{1}{k} t'_{a,k} \Gamma_{a,-k}
\quad , \quad
\gamma'=\Pi_V^{\parallel \Ker\hat B}\Gamma'.
\eeq

Denoting pairs $(a_j,k_j)$ as $i_j$, and $\bcycle_{a,k}=\frac{1}{k}\Gamma_{a,-k}$, we define for all $(g,n)$ such that $2g-2+n>0$
\beq
F_{g,n}[i_1,\dots,i_n] = \int_{\bcycle_{i_1}}\dots \int_{\bcycle_{i_n}} \omega_{g,n},
\eeq
\beq
F_{g,n}(t') = \sum_{i_1,\dots,i_n} F_{g,n}[i_1,\dots,i_n] t'_{i_1} \dots t'_{i_n}
 = \int_{\Gamma'}\dots\int_{\Gamma'} \omega_{g,n}= \int_{\gamma'}\dots\int_{\gamma'} \omega_{g,n}.
\eeq
and, as a formal $\hbar$ series
\beq
\ln Z(t') =  \sum_{(g,n), \ 2g-2+n>0} \frac{\hbar^{2g-2+n}}{n!} F_{g,n}(t') ,
\eeq
so that we have
\beq
\ln Z' = \hbar^{-1}\int_{\gamma'}\omega_{0,1} +\frac12 \int_{\gamma'}\int_{\gamma'}\omega_{0,2}+\ln Z(t').
\eeq
%Remark that
%\beq
%\int_{\gamma'}\omega_{0,1}
%= \sum_{i,j} t'_i t_j X_{i,j}
%\eeq
%\beq
%\Pi_V^{\parallel \Ker\hat B}\bcycle_i-\bcycle_i=\sum_j X_{i,j}\acycle_j.
%\eeq
Using 
\beq
\Delta \bcycle_{a,k} = \frac{k-r_a}{t_{a,r+1}}  \hat B(\Gamma_{a,1})\otimes  \bcycle_{a,k-r_a}
\qquad , \qquad
%\int_{\bcycle_{a,k}} \hat B(\Gamma_{a,j}) = \delta_{k,j},
\eeq
let us compute the action of $\Delta$ on $F_{g,n}$:
\bea
\Delta F_{g,n}[i_1,\dots,i_n]
&=& \sum_{a,k} F_{g,n+1}[i_1,\dots,i_n,(a,k)]  \hat B(\Gamma_{a,k}) \cr
&& + \sum_{j=1}^n  F_{g,n}[i_1,\dots, (a_j,k_j-r_{a_j}),\dots,i_n]  \frac{k_j-r_{a_j}}{t_{a_j,r_{a_j}+1}} \hat B(\Gamma_{a_j,1}) \cr
\eea
i.e.
\beq
\Delta F_{g,n}(t')
= \frac{1}{n+1} \sum_{a,k} \hat B(\Gamma_{a,k}) \frac{\partial}{\partial t_{a,k}} F_{g,n+1}(t')   
 + \sum_{a,k} k\frac{t'_{a,k+r}}{t_{a,r+1}} \hat B(\Gamma_{a_k,1})   \frac{\partial}{\partial t_{a,k}}  F_{g,n}(t') .
\eeq
Altogether that implies
\beq
\hbar\Delta \ln Z'
=   \Delta'  \ln Z   + \hat B(\gamma') +\int_{\Delta\gamma'} \omega_{0,1} + \hbar \int_{\gamma'}\int_{\Delta\gamma'} \omega_{0,2} 
\eeq
where we have defined
\beq
\Delta' = \sum_{a,k}  \left(  \hat B(\Gamma_{a,k}) +  \hbar k\frac{t'_{a,k+r}}{t_{a,r+1}} \hat B(\Gamma_{a_k,1}) \right) \frac{\partial}{\partial t'_{a,k}}.
\eeq
Moreover notice that
\beq
\hbar\Delta  Z=  \Delta' Z   - Z \frac{\hbar}{2} \int_{\gamma'}\int_{\gamma'}\omega_{0,3} .
\eeq

%Taking a second derivative
%\bea
%\hbar^2\Delta\Delta \ln Z'
%&=& \hbar [\Delta,\Delta']\ln Z + \hbar \Delta'\Delta \ln Z\cr
%&& + \hbar  \int_{\Delta\gamma'} \omega_{0,2}+ \hbar \int_{\gamma'} \omega_{0,3} \cr
%&& +\hbar \int_{\Delta\Delta\gamma'} \omega_{0,1}+\hbar\int_{\Delta\gamma'} \omega_{0,2} \cr
%&& + \hbar^2\int_{\Delta \gamma'}\int_{\Delta\gamma'} \omega_{0,2}  
%+ \hbar^2\int_{\gamma'}\int_{\Delta \Delta\gamma'} \omega_{0,2}  
%+ \hbar^2 \int_{\gamma'}\int_{\Delta\gamma'} \omega_{0,3} \cr
%%
%&=& \hbar \sum_{a,k}  \left(  \Delta\hat B(\Gamma_{a,k}) +  \hbar k\frac{t'_{a,k+r}}{t_{a,r+1}} \Delta\hat B(\Gamma_{a_k,1}) \right) \frac{\partial}{\partial t'_{a,k}} \ln Z  \cr
%&& + \Delta'\Delta' \ln Z -\frac12 \hbar \Delta' \int_{\gamma'}\int_{\gamma'}\omega_{0,3} \cr
%&& + \hbar  \int_{\Delta\gamma'} \omega_{0,2}+ \hbar \int_{\gamma'} \omega_{0,3} \cr
%&& +\hbar \int_{\Delta\Delta\gamma'} \omega_{0,1}+\hbar\int_{\Delta\gamma'} \omega_{0,2} \cr
%&& + \hbar^2\int_{\Delta \gamma'}\int_{\Delta\gamma'} \omega_{0,2}  
%+ \hbar^2\int_{\gamma'}\int_{\Delta \Delta\gamma'} \omega_{0,2}  
%+ \hbar^2 \int_{\gamma'}\int_{\Delta\gamma'} \omega_{0,3} \cr
%\eea
%

%
%Notice that the term $\hat B(\gamma')$ cancels with the one in \eqref{defoperatorK2}.
%It appears only within $K_2$ and we have defined
%\beq
% K_2(\hat B(\gamma'), .) = K_2^*(\gamma',.).
%\eeq
%Also
%\beq
%K_2(\hat B(\gamma'), \hat B(\gamma')) = \frac12 \int_{\gamma'}\int_{\gamma'}\omega_{0,3}
%= \frac12 \sum_{i,j} t'_i t'_j F_{0,3}[i,j,.].
%\eeq
In the end, theorem \ref{mainth2} can be rewritten as an operator which is quadratic in the $\partial/\partial t'_j$ and quadratic in the $t'_j$s, annihilating $Z(t')$.
It coincides with  Kontsevich-Soibelman \cite{KS17, ABCD17}.

More generally, we see that the equation \eqref{maineq2} (resp. \eqref{maineqk}) amounts to a order $\max r_a$ PDE with respect to times $t'_i$, and whose coefficients can themselves be polynomials of the $t'_i$ of the same order.

\subsubsection{Reminder KS method}

Let us rewrite KS proof. Notice that for all $(g,n)$ such that $2g-2+n>0$ we have
\beq
\omega_{g,n}(z_1,\dots,z_n) = \sum_{i_1,\dots,i_n} F_{g,n}[i_1,\dots,i_n] \ \hat B(\Gamma_{i_1})(z_1)\otimes \dots \otimes B(\Gamma_{i_n})(z_n)
\eeq
(indeed the difference between the LHS and RHS has all its $\bcycle_{a,k}$ integrals vanishing, which implies that it has no poles at branchpoints, but it also can't have other poles, therefore it must be a holomorphic form, and it is easy to see that the integral on $H_1(\curve,\CC)\cap \Ker\hat B)$ also vanishes, which implies that the difference is zero).
Let
\beq
C[i_1,i_2,i_3] = 2\int_{\bcycle_{i_1}} K_2(\hat B(\Gamma_{i_2}),\hat B(\Gamma_{i_3}))
\eeq
and
\beq
B[i_1,i_2,i_3] = 2\int_{z\in \bcycle_{i_3}} \left( \int_{\bcycle_{i_1}} K_2(\hat B(\Gamma_{i_2}),\omega_{0,2}(z,.)) \right)
\eeq
Then, topological recursion implies that for all $(g,n)\neq (0,0),(1,0),(0,1),(0,2)$ we have
\bea\label{eqTRFgnlocal}
2F_{g,n+1}[i_0,i_1,\dots,i_n]
&=& \sum_{i,j} C[i_0,i,j] \Big( F_{g-1,n+2}[i,j,i_1,\dots,i_n]   \cr
&& + \sum_{g_1+g_2=g, I_1\sqcup I_2=\{i_1,\dots,i_n\}}^{2g_i-2+n_i>0}    F_{g_1,1+|I_1|}[i,I_1] F_{g_2,1+|I_2|}[j,I_2] \Big)  \cr
&& + 2\sum_{k=1}^n \sum_j B[i_0,i_k,j] F_{g,n}[j,i_1,\dots,\widehat{i_k},\dots,i_n]
\eea
and in addition we have
\beq
F_{0,3}[i,j,k] = \int_{\bcycle_i}\int_{\bcycle_j}\int_{\bcycle_k} \omega_{0,3} = \frac12 A[i,j,k]
\eeq
\beq
F_{1,1}[i] = \int_{\bcycle_i} \omega_{1,1} = D[i].
\eeq
Then, \eqref{eqTRFgnlocal} can be written as the quadratic PDE
\beq
\Big(\hbar \frac{\partial}{\partial t'_i} -\hbar D[i] - \frac\hbar{2} \sum_{j,k} (A[i,j,k]t'_it'_j+2B[i,j,k]t'_j\hbar \frac{\partial}{\partial t'_k} + C[i,j,k]\hbar \frac{\partial}{\partial t'_j}\hbar \frac{\partial}{\partial t'_k}\Big) Z(t')=0
\eeq

\section{Conclusion}

For a spectral curve (either compact or local), 
the space of generalized cycles is the natural vector space on which the ABCD tensors of an Airy structure act.
The topological recursion is then equivalent to a differential operator annihilating a partition function.
This formalism makes rather easy the generalization to more than quadratic operators.
It should amount to \cite{borot2018higher} for W-algebra structures.

\section*{Aknowledgements}

This work is supported by the ERC Synergie Grant ERC-2018-SyG  810573 "ReNewQUantum".
It is also partly supported by the ANR grant Quantact : ANR-16-CE40-0017.
I wish to thank IHES for hosting me.
And I thank M. Kontsevich and Y. Soibelman for discussions.

\end{document}